\input{aipcheck}

\documentclass[
    ,final            % use final for the camera ready runs
  ]
  {aipproc}

\layoutstyle{6x9}

\begin{document}

\title{The effect of different opacity data and chemical element mixture on the Petersen diagram}

\classification{97.10.Sj
%<Replace this text with PACS numbers; choose from this list:
%                \texttt{http://www.aip..org/pacs/index.html}>
}

\author{P. Lenz}{
  address={Institute of Astronomy, University of Vienna, T\"urkenschanzstr. 17, 1180 Vienna, Austria}
}

\author{A. A. Pamyatnykh}{
  address={Institute of Astronomy, University of Vienna, T\"urkenschanzstr. 17, 1180 Vienna, Austria} 
  ,altaddress={Copernicus Astronomical Center, Polish Acad. Sci., Bartycka 18, 00-716 Warsaw, Poland\\and\\Institute of Astronomy, Russian Acad. Sci., Pyatnitskaya Str. 48, 109017 Moscow, Russia} % additional visiting address
%  ,altaddress={Institute of Astronomy, Russian Acad. Sci., Pyatnitskaya Str. 48, 109017 Moscow, Russia} % additional visiting address
}

\author{M. Breger}{
  address={Institute of Astronomy, University of Vienna, T\"urkenschanzstr. 17, 1180 Vienna, Austria}
}

\begin{abstract}
The Petersen diagram is a frequently used tool to constrain model parameters such as metallicity of radial double-mode pulsators. In this diagram the period ratio of the radial first overtone to the fundamental mode, $\Pi_1/\Pi_0$,  is plotted against the period of the fundamental mode. The period ratio is sensitive to the chemical composition as well as to the rotational velocity of a star. In the present study we compute stellar pulsation models to demonstrate the sensitivity of the radial period ratio to the opacity data (OPAL and OP tables)  and we also examine the effect of different relative abundances of heavy elements. We conclude that the comparison with observed period ratios could be used successfully to test the opacity data.
\end{abstract}

\maketitle

%%%%%%%%%%%%%%%%%%%%%%%%%%%%%%%%%%%%%%%%%%%%
%% MAINMATTER
%%%%%%%%%%%%%%%%%%%%%%%%%%%%%%%%%%%%%%%%%%%%

\section{Introduction}

Petersen diagrams are a widely used tool in asteroseismology. These diagrams were introduced by \cite{petersen:1973} to study double-mode cepheids. It was shown by \cite{suarez:2006} that Petersen diagrams are sensitive not only to metallicity but also to stellar rotation.  In this investigation we test the influence of the choice of opacity data and metal mixtures on the period ratio. Hitherto these effects have mainly been taken into account for studies on the instability domains of stars in the HR diagram  (see, for example, Pamyatnykh \& Ziomek 2007\cite{pamyatnykh:2007} and Miglio et al. 2007\cite{miglio:2007}). 
Our examination is based on models of the $\delta$~Scuti star 44~Tau. In addition to a number of nonradial modes this star pulsates in the radial fundamental (6.8980 c/d) and the first overtone mode (8.9606 c/d).  A detailed frequency analysis of the observational data of 44~Tau has been published by \cite{antoci:2007}. The presence of two radial modes with known radial order reduces the number of possible models significantly. For a certain set of input parameters, such as X, Z, $\alpha_{\rm MLT}$, $\alpha_{\rm ov}$ and v$_{\rm rot}$, only one model with a specific mass fits both the fundamental and first overtone frequency. \cite{zima:2007} confirmed that the star is an intrinsically slow rotator with a measured rotational velocity of 1-5 km/s.  They also determined the photospheric element abundances  and  found  no  significant  deviations  from  the  solar  values.  Due  to  these  strong  constraints  44~Tau is a very good target to examine the influence of  different opacity tables and element mixtures. 
For our study we are using the same codes as described in \cite{olech:2005}. Only nonrotating models are considered.

\section{Effect of Opacity Tables}

\begin{figure}
  \includegraphics[height=.25\textheight]{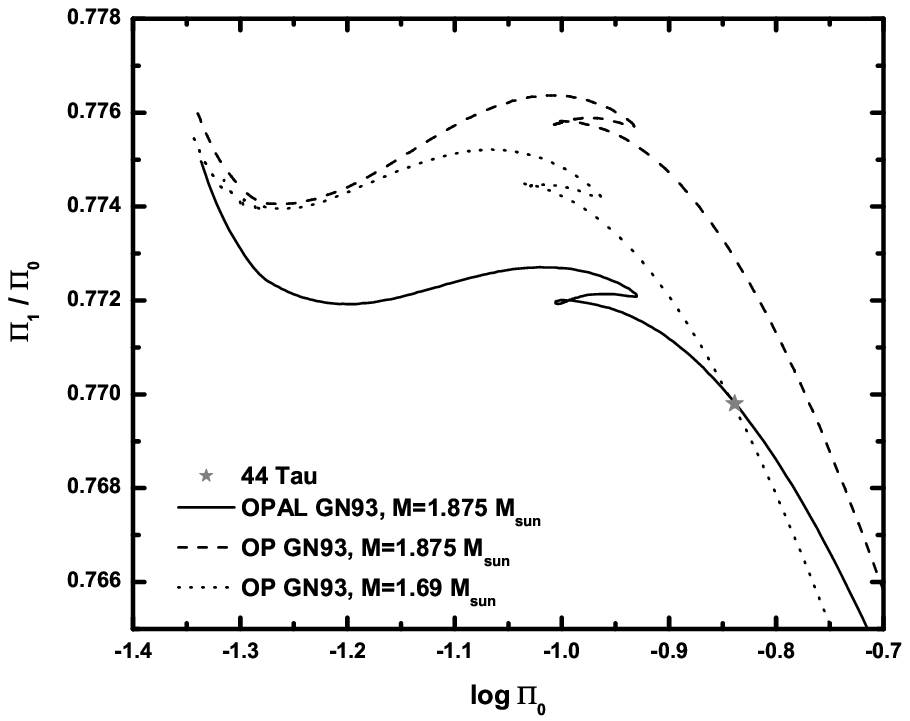}
  \includegraphics[height=.25\textheight]{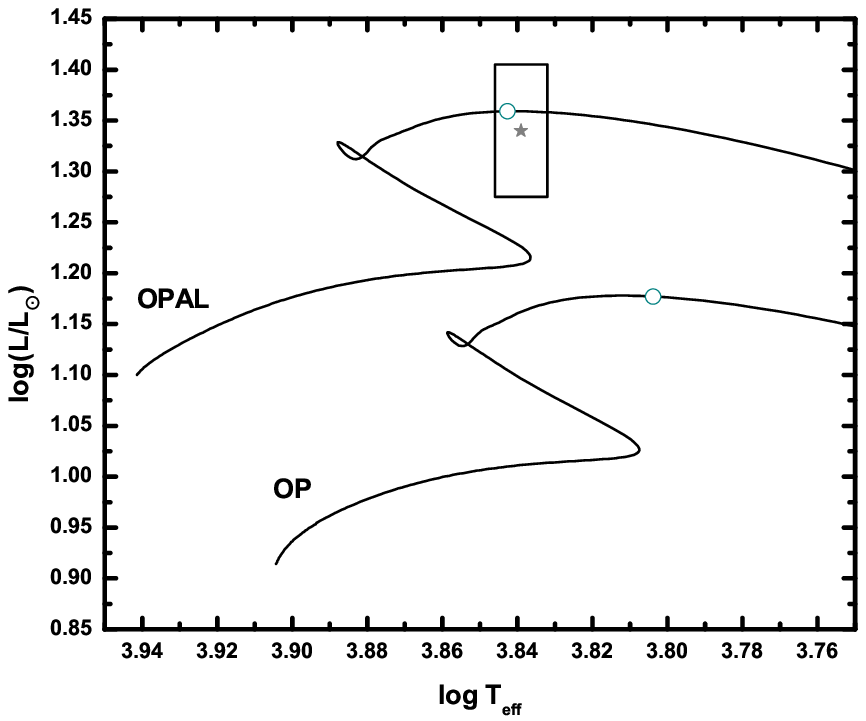}
  \caption{Left panel: Petersen diagram comparing the results obtained with the  OPAL and OP opacity tables.  The grey asterisk corresponds to the position of the slowly rotating $\delta$~Scuti star 44~Tau. Right panel: Position of the models that fit the observed radial modes of 44~Tau in the HR diagram. The box indicates the expected position derived from  photometric measurements.}
  \label{fig:pethrd}
\end{figure}

The latest version of the OPAL opacities dates back to 1996 (Iglesias \& Rogers 1996). The OP opacities have recently been updated (Seaton 2005). \cite{badnell:2005} compared OP and OPAL opacities and concluded that the new OP tables are much closer to the OPAL tables. They found that the agreement between OPAL and OP opacities is mainly within 5-10\%. 
Using standard chemical composition and the GN93 metal mixture (Grevesse \& Noels 1993), we compared the results obtained with OPAL and OP tables for a 1.875 M$_{\odot}$ model. As can be seen in the Petersen diagram in Fig.~\ref{fig:pethrd}, the difference is indeed significant. For OP opacities the period ratio $\Pi_1/\Pi_0$ is predicted to be higher. Consequently, to obtain a good fit of the observed radial modes the mass has to be reduced to 1.69 M$_{\odot}$. In the right panel of Fig.~\ref{fig:pethrd} the position of the fitted models in the HR diagram is shown.  While the OPAL model is located within the error box from photometric observations, the OP model is clearly too cool and too faint. The parameters of the fitted models are given in Table~\ref{table}.

\subsection{Comparison of OPAL and OP opacities}

\begin{figure}
  \includegraphics[height=.25\textheight]{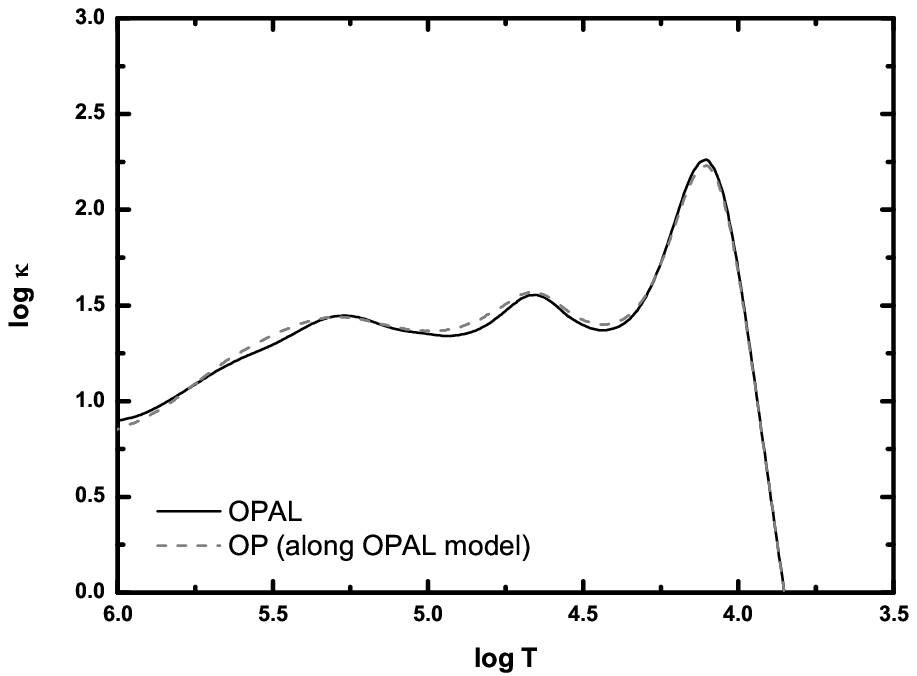}
  \includegraphics[height=.25\textheight]{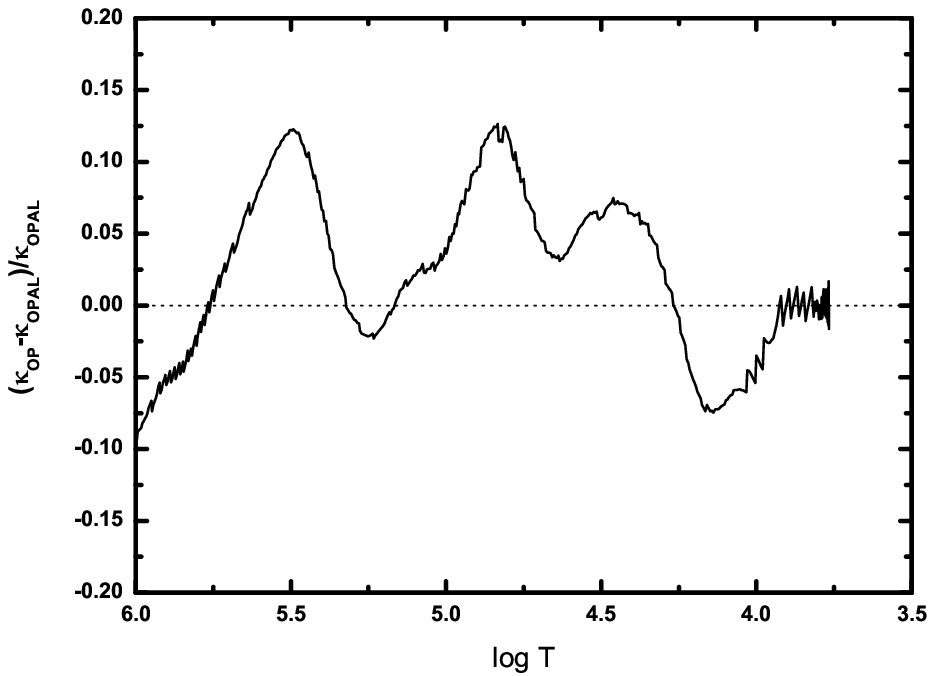}
  \caption{Left panel: Rosseland-mean opacities obtained from OP tables using the temperature and density distribution from  the OPAL  model. The right panel shows the relative differences in opacities between these models.}
  \label{fig:opac1}
\end{figure}

\begin{figure}
  \includegraphics[height=.25\textheight]{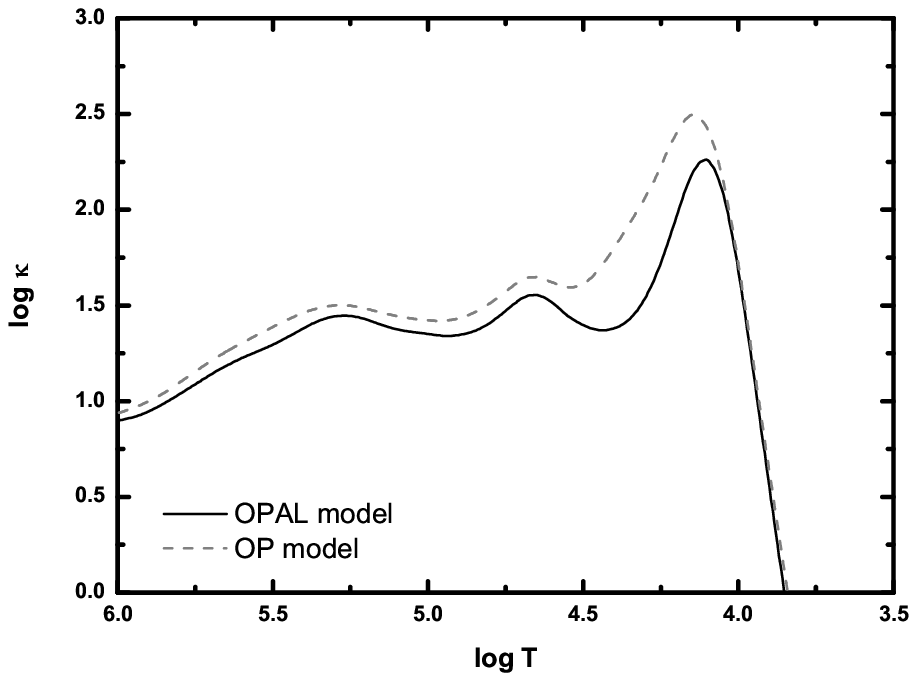}
  \includegraphics[height=.25\textheight]{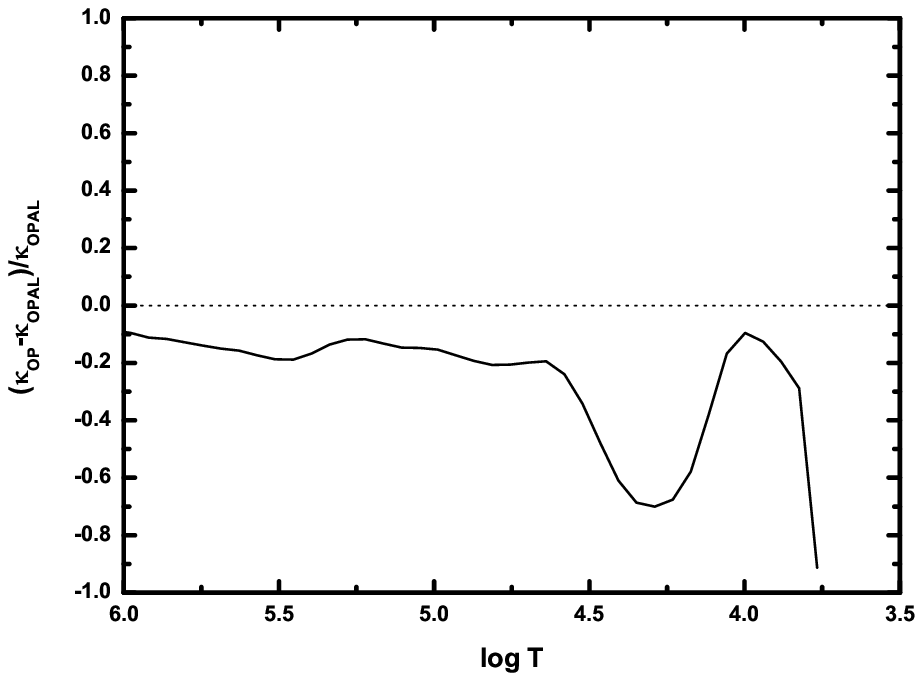}
  \caption{Left panel: Rosseland-mean opacities  for the OPAL  vs. OP model of 44~Tau. Both models fit the observed radial frequencies. The right panel shows the relative differences in opacities between these models.}
  \label{fig:opac2}
\end{figure}

To examine the cause for the differences in the Petersen diagram in more detail we compared the Rosseland-mean opacities, $\kappa$, of the stellar models. Using the temperature and density distribution inside the OPAL model of 44~Tau, we computed the corresponding opacities from the OP tables. As can be seen in Fig.~\ref{fig:opac1} the results are similar. The relative differences do not exceed 15 \%. A larger deviation occurs if we compare the opacities of the two fitted 44~Tau models obtained with OP and OPAL tables. While both models fit the observed radial modes, global parameters such as the effective temperature are different. Fig.~\ref{fig:opac2} shows that the difference is generally at the level of 20\% or larger.


\begin{table}
  \begin{tabular}{llllllllll}
\hline
  & \tablehead{1}{l}{b}{Opacity}
  & \tablehead{1}{l}{b}{Element mixture}
  & \tablehead{1}{l}{b}{X}
  & \tablehead{1}{l}{b}{Z}
  & \tablehead{1}{l}{b}{M/M$_{\odot}$} 
  & \tablehead{1}{l}{b}{$\log T_{\rm eff}$}
  & \tablehead{1}{l}{b}{$\log L$}
  & \tablehead{1}{l}{b}{$\log g$} 
  & \tablehead{1}{l}{b}{Age [Myr]} \\
\hline
 & OPAL & GN93 & 0.70 & 0.02  & 1.875 & 3.8422 & 1.3601 & 3.6712 & 1120 \\
 & OP   & GN93 & 0.70 & 0.02  & 1.695 & 3.8052 & 1.1822 & 3.6571 & 1520 \\
 & OPAL & A04  & 0.70 & 0.02  & 1.860 & 3.8313 & 1.3070 & 3.6767 & 1250 \\
 & OPAL & A04  & 0.74 & 0.012 & 1.805 & 3.8303 & 1.3035 & 3.6635 & 1330 \\
\hline
  \end{tabular}
  \caption{Parameters of the models for which the predicted  radial fundamental and first overtone frequencies were fitted to the observed values of 44~Tau. For all models inefficient convection ($\alpha_{\rm MLT} = 0.2$) and no overshooting from the convective core were assumed.}
  \label{table}
\end{table}

\section{Effect of Chemical Element Mixture}

\begin{figure}
  \includegraphics[height=.25\textheight]{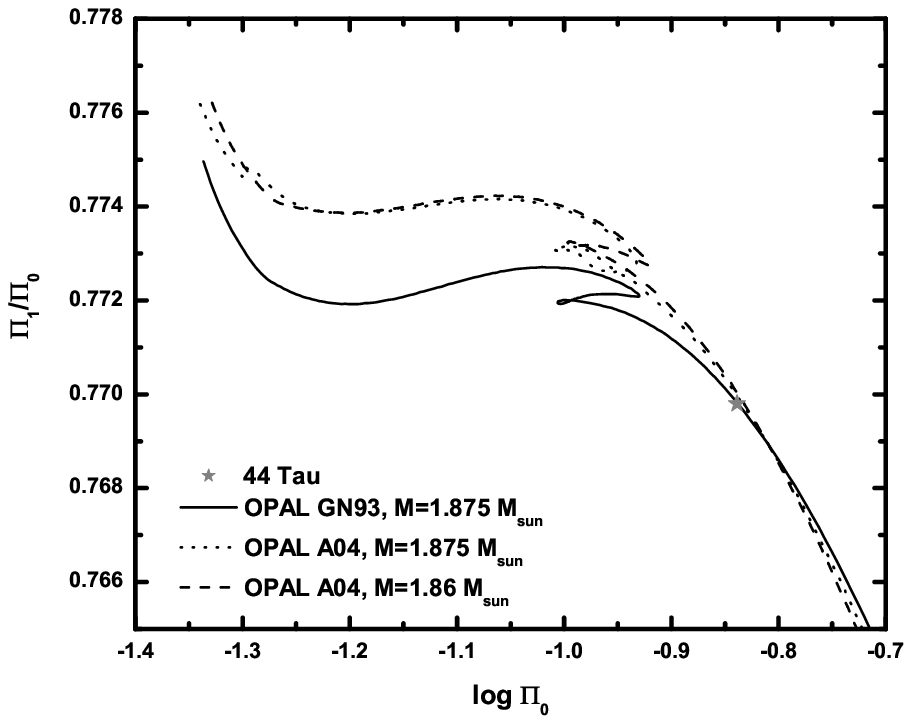}
  \includegraphics[height=.25\textheight]{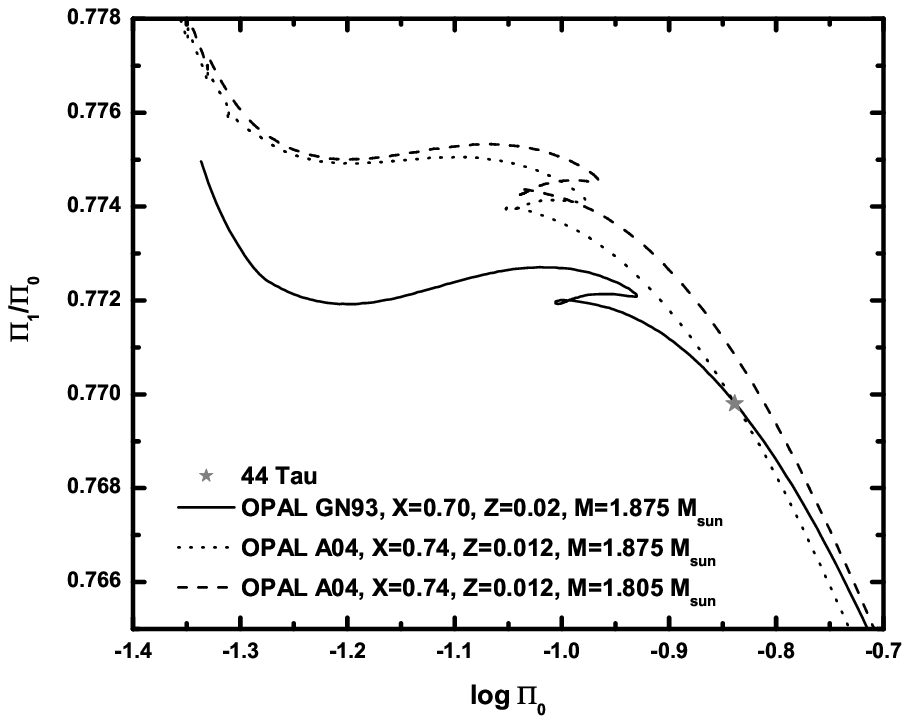}
  \caption{Left panel: Effect of the choice of element mixture on the Petersen diagram . All models were computed for  X=0.70 and Z=0.02. Right panel: Comparison of the standard  GN93 model with  the model computed for the new values for the solar composition.}
  \label{fig:petersen}
\end{figure}

\begin{figure}
  \includegraphics[height=.25\textheight]{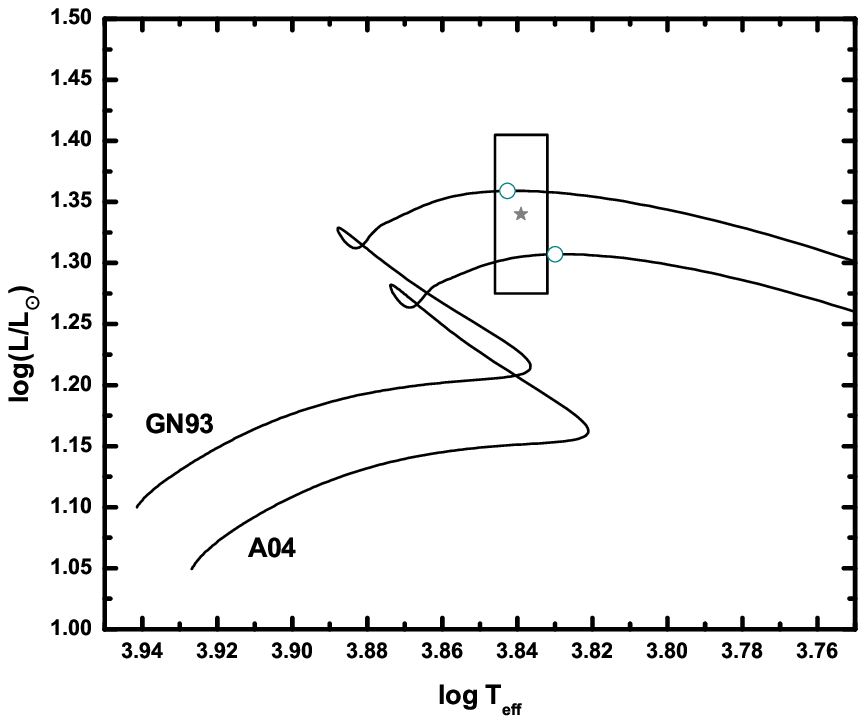}
  \includegraphics[height=.25\textheight]{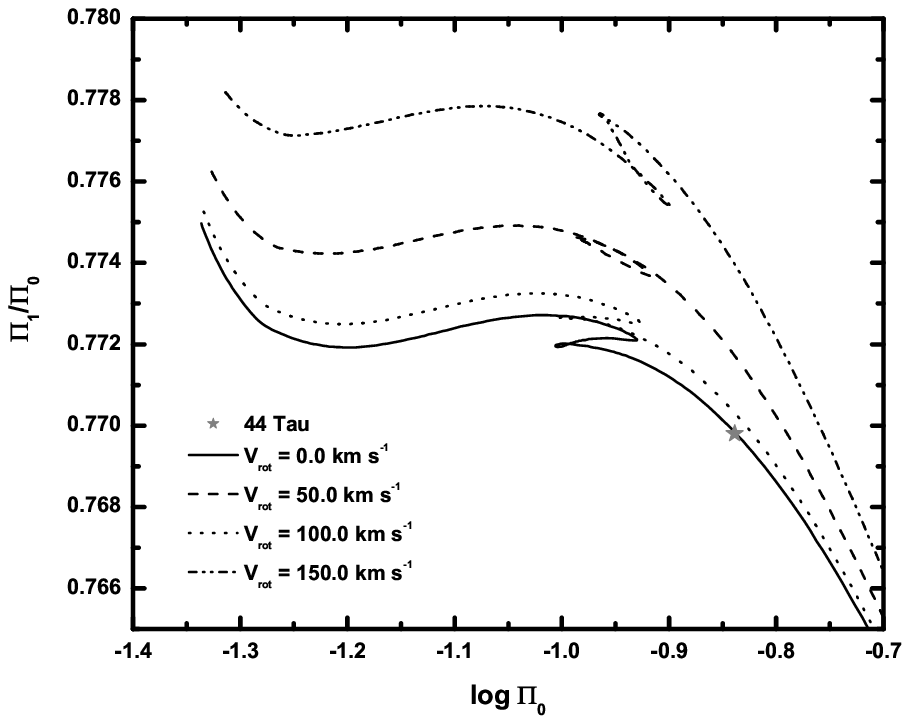}
  \caption{Left panel: Position of the fitted models shown in the right panel of Fig.~\ref{fig:petersen} in the HR diagram. Right panel: The effect of rotation on the Petersen diagram.}
  \label{fig:mixrot}
\end{figure}

A recent analysis of the solar spectrum by Asplund et al. (2004, 2005) gave rise to a significant downward revision of the solar C, N, O and Ne abundances by 40-60\% and to a moderate downward revision of Fe-group abundances by 10-25\%. The solar composition corresponding to the new metal mixture A04 has been determined to be X=0.74, Z=0.0122. These results have destroyed the good agreement between the standard solar model and the helioseismic model. 
The influence of the choice between the GN93 and A04 metal mixture on the Petersen diagram is shown in the left panel of Fig.~\ref{fig:petersen}. The models were computed using  the OPAL opacity tables and assuming X=0.70 and Z=0.02. In the right panel of Fig.~\ref{fig:petersen} the combined effect of element mixture and the new solar X and Z value is displayed. The deviation in the Petersen diagram is significant as well. However, it is possible to fit the observed radial modes close to the photometric error box. The comparison with the effect of rotation on the Petersen diagram (Fig.~\ref{fig:mixrot}, right panel) shows that the deviations are approximately at the same level. In the computations for this diagram we assumed uniform rotation and conservation of global angular momentum during the stellar evolution. The pure effect of changing the metallicity from Z=0.02 to 0.0122 increases the period ratio by a value of almost 0.002. 

\section{Conclusions}

In this work we tested the influence of different opacity tables and chemical element mixtures on the Petersen diagram based on models of the $\delta$~Scuti star 44~Tau. We found that the choice between OPAL and OP opacities has a significant effect on the Petersen diagram. The seismic model computed with the OP tables does not match the photometrically derived position of 44~Tau in the HRD: the OP model is clearly too cool and too faint. The OPAL opacities are preferable to explain the observed radial modes and fundamental parameters in the case of 44~Tau. 
The new solar heavy element mixture by Asplund et al. (2004, 2005) and the new solar X and Z values (X=0.74, Z=0.0122) have a smaller effect on the Petersen diagram.  We do not find significant indications in favor of the GN93 or the A04  element mixture. The differences caused by the use of different opacity data and metal mixtures are approximately of the same size as those caused by rotational effects (at v$_{\rm rot} \approx$ 100 km/s) and should also be considered when using the Petersen diagram as a diagnostic tool.

%%%%%%%%%%%%%%%%%%%%%%%%%%%%%%%%%%%%%%%%%%%%%%%%
%% BACKMATTER
%%%%%%%%%%%%%%%%%%%%%%%%%%%%%%%%%%%%%%%%%%%%%%%%

\begin{theacknowledgments}
This  work has been supported by the Austrian Fonds zur F\"orderung der wissenschaftlichen Forschung. AAP acknowledges partial financial support from the Polish MNiSW grant No. 1 P03D  021 28.
\end{theacknowledgments}



\bibliographystyle{mn2e}   % if natbib is available

%\bibliography{lenz}

%  \typeout{******************************************}
%  \typeout{** Please run "bibtex \jobname" to optain}
%  \typeout{** the bibliography and then re-run LaTeX}
%  \typeout{** twice to fix the references!}
%  \typeout{******************************************}

%%%%%%%%%%%%%%%%%%%%%%%%%%%%%%%%%%%%%%%%%%%
%% The following lines show an example how to produce a bibliography
%% without the help of the BibTeX program. This could be used instead
%% of the above.
%%%%%%%%%%%%%%%%%%%%%%%%%%%%%%%%%%%%%%%%%%%

\end{document}